\newif\ifreview
\reviewfalse        

\ifreview
  \documentclass[preprint,1p,times]{elsarticle}
\else
  \documentclass[5p,times]{elsarticle}
\fi

\ifreview
  \usepackage{lineno}
\fi
\usepackage{graphicx}
\usepackage{amsmath}
\usepackage{amssymb}
\usepackage{booktabs}
\usepackage{array}
\usepackage{multirow}
\usepackage{subcaption}
\usepackage{siunitx}
\usepackage{xcolor}
\usepackage{float}

\journal{Nuclear Instruments and Methods in Physics Research A}

\begin{document}

\ifreview
\linenumbers
\fi

\begin{frontmatter}

\title{Water Cherenkov Detectors in Precision Agriculture: A Novel Approach for High-Resolution Soil Moisture Monitoring}

\author[1,2]{Christian Sarmiento-Cano}
\author[2]{Jaime Betancour}
\author[3,4]{Alejandro Núñez Selin\corref{cor1}}
\author[2]{Luigui Miranda-Leuro}
\author[3]{Iván Sidelnik}
\author[5]{Hernán Asorey}
\author[2,6]{Luis A. Núñez}

\address[1]{Departamento de Ciencias Básicas, Universidad Autónoma de Bucaramanga, Bucaramanga, Colombia}
\address[2]{Escuela de Física, Universidad Industrial de Santander, Bucaramanga, Colombia}
\address[3]{Departamento de Física de Neutrones, Centro Atómico Bariloche, San Carlos de Bariloche, Argentina}
\address[4]{Instituto Balseiro, CNEA, San Carlos de Bariloche, Argentina}
\address[5]{piensas.xyz, Las Rozas Innova, Las Rozas de Madrid, Spain}
\address[6]{Depatamento de Física, Universidad de Los Andes, Mérida, Venezuela}

\begin{abstract}
Water Cherenkov Detectors (WCDs), traditionally employed in cosmic-ray detection, are repurposed here for precision soil moisture monitoring using cosmic-ray neutron sensing. This approach offers advantages over conventional neutron probes, including enhanced sensitivity to low moisture levels and the ability to cover larger soil volumes without subsurface intrusion.
This study evaluates the feasibility of WCDs for agricultural neutron hydrometry, addressing challenges such as background suppression and data interpretation in heterogeneous soils. We present experimental results from controlled wet and dry soil-condition emulations, alongside Monte Carlo simulations using Geant4 with an atmospheric neutron spectrum to correlate signal variation with soil-moisture differences.
By bridging particle physics and agronomy, WCDs could advance soil moisture monitoring, offering a non-invasive, scalable, and accurate alternative for optimizing agricultural water use. Preliminary findings suggest a transformative potential for sustainable farming, though further research is needed to enhance cost-efficiency and adaptability to diverse soil types.

\end{abstract}


\end{frontmatter}

\section{Introduction}

Monitoring soil moisture is fundamental for the sustainable management of water resources in agriculture, where precision irrigation can significantly enhance crop yield and conserve water. Cosmic-ray neutron sensing (CRNS) has emerged as a powerful, non-invasive technique for estimating soil water content at the hectare scale, based on the inverse correlation between the detected epithermal-neutron flux and the hydrogen content of soil moisture. While traditional CRNS networks, such as COSMOS, rely on high-pressure \(^3\)He proportional counters — known for their efficiency — they face critical limitations due to the global shortage and rising cost of \(^3\)He gas~\cite{sachetti_3he_free_2015}, hindering their widespread adoption, particularly in developing agricultural regions.

This supply challenge has prompted research into alternative neutron detection technologies. Among them, water Cherenkov detectors (WCDs) present a compelling solution. Originally developed for large-scale cosmic-ray observatories (e.g., the Pierre Auger Observatory~\cite{auger_observatory_2015}), WCDs can detect neutrons indirectly via the secondary particles produced in nuclear interactions. Specifically, they are sensitive to the 2.22~MeV gamma rays emitted from neutron capture on hydrogen and can utilize dopants like sodium chloride (NaCl) to enhance thermal neutron capture on chlorine-35, producing a distinctive cascade of higher-energy gammas~\cite{sidelnik2020enhancing}. This dual detection mechanism, coupled with the use of low-cost, non-toxic, and readily available materials (water and salt), makes WCDs a scalable and economically viable alternative to \(^3\)He-based systems.

In this context, this work evaluates the feasibility of a NaCl-doped WCD as a novel sensor for agricultural soil moisture monitoring. We employ a combined approach of detailed Monte Carlo simulations using the Geant4 toolkit within the MEIGA framework ~\cite{Taboada2022} and experimental validation with a neutron source. The results aim to bridge particle physics instrumentation and agronomic practice by proposing a robust, accessible technology that could democratize large-scale, precision soil moisture monitoring for sustainable farming.

\section{Experimental Setup}

\subsection{Cosmic Neutron Flux Simulation}

To simulate the production and propagation of cosmic-ray-induced neutrons in the atmosphere, we employed the ARTI simulation toolkit~\cite{sarmiento_arti_2022}, developed by the Latin American Giant Observatory (LAGO) collaboration~\cite{LAGO_rep}. ARTI is a modular framework designed to quantify the background flux of secondary particles from galactic cosmic rays (GCRs). It integrates three core components: \emph{MAGCOS} for calculating the geomagnetic rigidity cutoff at a given location, \emph{CORSIKA} for simulating extensive air showers initiated by high-energy primaries, and \emph{Geant4} for modeling the low-energy particle transport and detector response. This chained approach allows for a complete description of the neutron flux from its production in the upper atmosphere to its detection at ground level.

\begin{figure}[h!]
\centering
\begin{subfigure}{0.75\linewidth}
\includegraphics[width=\linewidth]{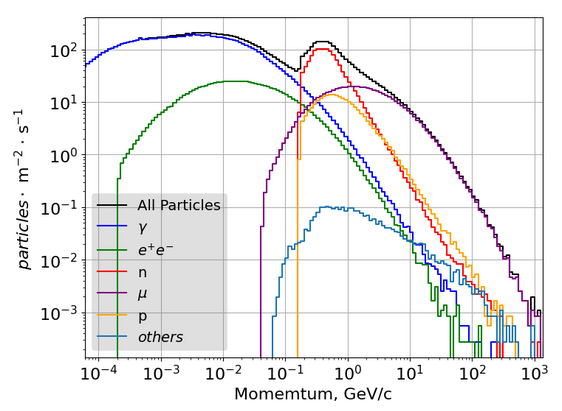}
\end{subfigure}
\hfill
\begin{subfigure}{0.80\linewidth}
\includegraphics[width=\linewidth]{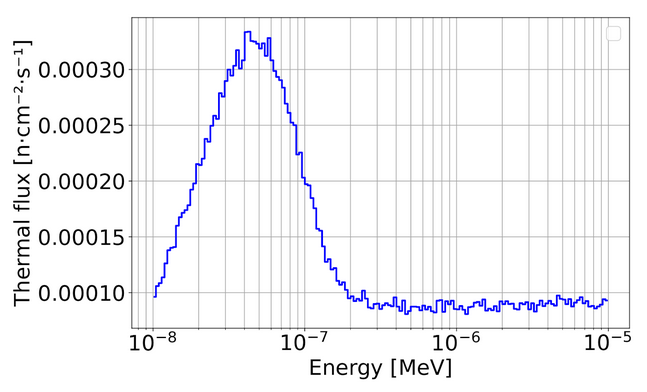}
\end{subfigure}
\caption{(top) The momentum spectra of secondary particles at ground level in Bariloche, Argentina, simulated using the ARTI framework. A cutoff at approximately 200 MeV/c is observed in the neutron spectrum, corresponding to the simulation's low-energy threshold. (down) The spectrum of thermal neutrons is simulated in MEIGA by means of the injection of the total flux of secondary particles. The neutrons were injected directly into the WCD.}
\label{fig:spectrum_buca}
\end{figure}

For this study, ARTI was configured for the geographic coordinates of Bariloche, Argentina (893 m a.s.l.) to generate the momentum spectrum of secondary particles at ground level. The resulting differential flux, decomposed into its neutron, gamma, electron, muon, and proton components, is shown in Figure~\ref{fig:spectrum_buca} (Top). A characteristic low-energy cutoff for neutrons around 20 MeV/c is observed, corresponding to the simulation threshold of the CORSIKA module. To recover the crucial thermal and epithermal neutron component ($<1$ keV), we performed complementary low-energy neutron transport simulations with MEIGA (Geant4).

The atmospheric model for the simulations was implemented as a stratified, dry-air volume (G4\_AIR) composed of 78\% N\textsubscript{2}, 21\% O\textsubscript{2}, and 1\% Ar, in hydrostatic equilibrium, following the U.S. Standard Atmosphere. The lowermost 2 km of the atmosphere was divided into ten horizontal layers of equal thickness. The \texttt{QGSP\_BERT\_HP} physics list was selected for its high-precision (\emph{HP}) treatment of neutron interactions below 20 MeV, using evaluated nuclear data libraries (ENDF/B-VII). This setup accurately models the elastic scattering, moderation, and capture processes that shape the thermal neutron flux emerging from the soil.

\subsection{Water Cherenkov Detector Model and Configurations}

The response of the Water Cherenkov Detector to these neutron fields was simulated using the MEIGA framework \cite{Taboada2022}, which interfaces with Geant4 for particle tracking and detector response. The WCD was modeled as a stainless-steel cylinder with a height of 133~cm and a diameter of 96~cm ($V \approx 0.96$~m\textsuperscript{3}), and a wall thickness of 0.5~mm. The inner walls were lined with a 0.12~mm layer of Tyvek, whose optical properties were characterized using a unified surface model. A hemispherical volume (10.16~cm radius) at the top center of the tank simulated the photocathode of an 8-inch Hamamatsu R5912 photomultiplier tube (PMT). The quantum efficiency (QE) of the PMT was implemented as a wavelength-dependent filter, accepting photons primarily in the 300--600~nm range with a peak efficiency of 25\% at 400~nm.

The active volume of the detector was defined with four distinct media:
\begin{enumerate}
    \item[(i)] Pure water ($\rho = 0.9982$~g/cm\textsuperscript{3} at 20$^\circ$C).
    \item[(ii)] Water with 2.5\% mass concentration of NaCl ($\rho = 1.0179$~g/cm\textsuperscript{3}).
    \item[(iii)] Water with 5.0\% mass concentration of NaCl ($\rho = 1.0342$~g/cm\textsuperscript{3}).
    \item[(iv)] Water with 10.0\% mass concentration of NaCl ($\rho = 1.0707$~g/cm\textsuperscript{3}).
\end{enumerate}

The optical properties of each medium, including the wavelength-dependent index of refraction, absorption length, and Rayleigh scattering length, were adjusted according to their salinity. The primary physics list for these simulations was \texttt{QGSP\_BERT\_HP}, supplemented by the \texttt{OpticalPhysics} module for the generation and transport of Cherenkov photons.

To study the detector's response to soil-moisture-dependent neutrons, we implemented a ground model in MEIGA to generate thermal-neutron albedo. A volume of dry Colombian soil, with a density of 2.7~g/cm\textsuperscript{3} and an elemental composition based on local samples (49\% O, 33\% Si, 7.1\% Al, with minor fractions of Fe, Ca, K, and H), was placed within the atmospheric volume. The atmospheric neutron flux simulated by ARTI was injected into this soil volume. The resulting thermalized neutrons that back-scatter into the air—the \emph{albedo flux}—were then recorded as a source plane 20 cm above the ground. This albedo spectrum, representative of dry soil conditions in Bucaramanga, was used as the primary input for detector performance studies (Figure~\ref{fig:spectrum_buca}, Down).

\subsection{Experimental Validation with a Neutron Source}

To validate the Geant4 detector model, we performed a controlled experiment using a physical WCD prototype. The setup consisted of a cylindrical stainless-steel tank with a capacity of 500 L, filled with the medium under test (pure water, 2.5\% NaCl, or 10\% NaCl). The detector was irradiated by a shielded $^{241}$AmBe neutron source (activity $\sim$1 Ci). A 10~cm-thick lead block was placed between the source and the detector to attenuate the direct 4.43~MeV gamma rays from the source, thereby isolating the neutron-induced signal.

The data acquisition system was based on a Red Pitaya board sampling at 125~MHz (8~ns intervals). For each system-triggering event, a 32-sample waveform was recorded. The integrated charge of each pulse (in ADC units) was calculated after baseline subtraction. Measurements were taken for 5-minute intervals for each detector medium configuration, followed by a background measurement with the source removed. The net neutron spectrum for each configuration was obtained by bin-by-bin subtraction of the background.

\subsection{Results}

\begin{figure}[h!]
\centering
\begin{subfigure}{0.95\linewidth}
\includegraphics[width=\linewidth]{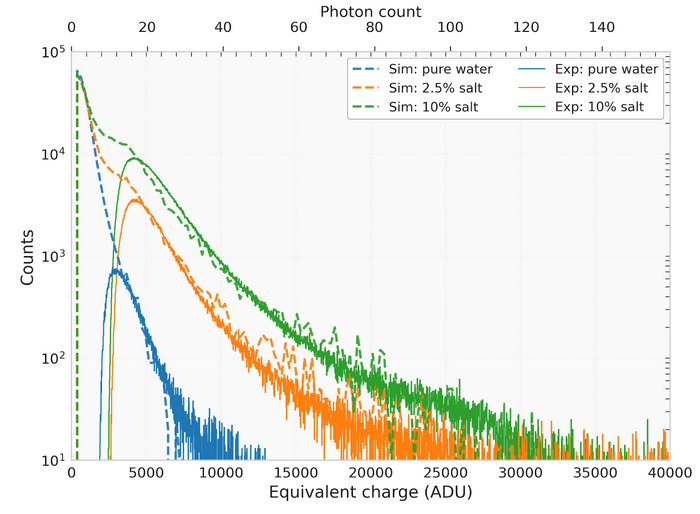}
\end{subfigure}
\hfill
\begin{subfigure}{0.60\linewidth}
\includegraphics[width=\linewidth]{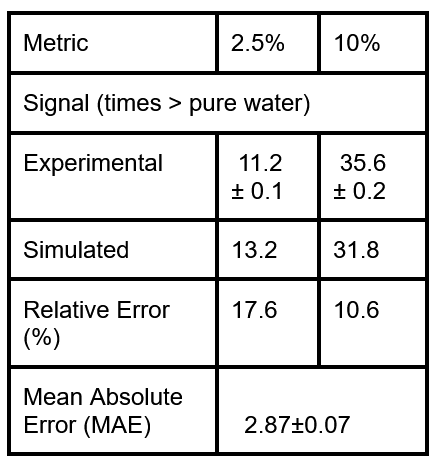}
\end{subfigure}
\caption{(top) Comparison between the experimental spectrum obtained using a \(^{241}\)AmBe source shielded with lead and the simulated Cherenkov photon spectra produced by prompt gamma rays from neutron captures in each medium, as they interact with atomic electrons. The experimental spectra exhibit an artificial low-energy cutoff due to the trigger threshold applied during data acquisition. (down) Comparison of experimental and simulated signal enhancement ratios (relative to pure water) for 2.5\% and 10\% NaCl solutions. Reported uncertainties in experimental values originate from Poisson counting statistics. The MAE incorporates propagated experimental errors.}
\label{datasim}
\end{figure}

 In water Cherenkov detectors, the medium naturally thermalizes incident neutrons through elastic scattering  with  \(^1\)H nuclei, while simultaneously serving as the Cherenkov radiator. The detectable signal is determined
 solely by the characteristic gamma emissions from neutron capture reactions either  \(^1\)H(n,$\gamma$) \(^2\)H (2.2 MeV) or  \(^35\)Cl(n,$\gamma$) \(^36\)Cl (1-8 MeV). Since these nuclear transitions depend only on the absorber's properties, the final Cherenkov signal becomes independent of the neutrons' initial energies. Figure 2 compares the simulated Cherenkov photon spectra with experimental charge spectra from field measurements. The signal enhancement ratios (relative to pure water) were measured at 11.2 $\pm$ 0.1 for the 2.5\% NaCl solution and 35.6 $\pm$ 0.2 for the 10\% NaCl solution, while simulations predicted ratios of 13.2 and 31.8, respectively. Model performance was quantified using two metrics: (1) a mean absolute error (MAE) of 2.87 ± 0.07, calculated as $\frac{1}{n} \sum_{i=1}^{n} |E_i - S_i|$ where E$_{i}$ and S$_{i}$ represent experimental and simulated obtained ratios, and (2) relative errors of 17.6\% (2.5\% NaCl) and 10.6\% (10\% NaCl), determined through $\frac{|E_i - S_i|}{E_i}$x$100\%$. The experimental uncertainties originate from Poisson counting statistics in the detector measurements. The experimental data and simulations are in good agreement, indicating that the physical processes involved in photon production and detection, such as neutron-induced gamma emissions and neutron backscattering, are well represented in the model (figure \ref{datasim}). This confirms that water Cherenkov detectors can be employed as a non-invasive tool to assess soil moisture, making them a promising candidate for applications in precision agriculture and environmental monitoring. Further details regarding the simulation pipeline and comparative analysis are available in the accompanying preprint \cite{Betancourt2025}

\section*{Acknowledgments}
The authors acknowledge co-funding from the Programa Iberoamericano de Ciencia y Tecnología para el Desarrollo (CYTED) through the LAGO-INDICA network (Project 524RT0159-LAGO-INDICA: Infraestructura digital de ciencia abierta). This work was partially supported by the CLAF-HECAP Programme through a research mobility scholarship. We also acknowledge the computational support from the Universidad Industrial de Santander (SC3UIS) High Performance and Scientific Computing Centre.

\end{document}